# An Accelerometer Based Instrumentation of the Golf Club:

## Measurement and Signal Analysis


Robert D. Grober
Department of Applied Physics
Yale University
New Haven, CT  06520


December 16, 2009


Two accelerometers are used to measure the motion of the golf club.  The accelerometers are mounted in the shaft of the golf club.  Each measures the acceleration along the axis of the shaft of the golf club.  Interpreting the measurement with the context of the double pendulum model of the golf swing, it is useful to resolve the resulting signals into differential and common mode components.  The differential mode is a measure of the rotational kinetic energy of the golf club, and this can be used to understand the tempo, rhythm, and timing of the golf swing.  The common mode measurement is related to the motion of the hands.  It is shown that both components can be used to recover the motion of the swing within the context of the double pendulum model of the golf swing.


Introduction

The use of electronics in the shaft or club head of a golf club has been the subject of considerable past work. Modern implementations offer a large number of sensors and computational power concealed within the shaft [1]. Over time, the tendency has been to make ever more sophisticated measurements in an effort to obtain increasingly detailed understanding of the golf swing. This paper describes a relatively simple measurement which yields a remarkably robust data set.

The measurement system consists of two accelerometers mounted in the shaft of a golf club with the direction of sensitivity oriented along the axis of the shaft. One accelerometer is located under the grip, preferably at a point between the two hands. The other is located further down the shaft. Interpretation of this data is based on the double pendulum model of the golf swing [2]. Within this formalism, the two accelerometers yield a common mode signal and a differential mode signal. The differential mode signal is proportional to the rotational kinetic energy of the golf club. This simple and robust signal provides insight into the tempo, timing and rhythm of the golf swing. Its simplicity enables real-time biofeedback [3]. The common mode signal measures the motion of the hands. It is shown that these two signals can be used to quantitatively determine the entire motion of the double pendulum throughout the golf swing.

Measurement Details

The wireless measurement system described below allows data collection in real-time, meaning the data is collected while a golfer swings the club. The measurement is based on the use of two accelerometers. Analog Device's ADXL193 (120 $g$) and

ADXL78 (50 *g*) single axis accelerometers are mounted in the shaft of a golf club with the sensing axis oriented along the axis of the shaft. The ADXL193 is mounted towards the club head while the ADXL78 is mounted under the grip of the club. The accelerometers produce a linear, analog output, which is digitized using the Microchip MCP3201, a 12-bit analog to digital converter. Data acquisition is mediated by a Microchip 16F873A microcontroller. The resulting data is communicated from the club to a base station using a Chipcon CC1000 wireless transceiver pair operating in the 915 MHz communications band. The base station interfaces to a computer, enabling data storage and signal analysis. The resulting data rate for the entire wireless system is 4.4 ms/cycle, each cycle yielding data from both accelerometers. While this under samples the 400 Hz low pass filter built into the output stage of the accelerometers, this update rate is sufficient for the time scales appropriate to all aspects of the golf swing, except perhaps in the few milliseconds near impact.

An example of the measured noise voltage for the 50 *g* accelerometer is shown in Fig. 1. When operated from a 5V power supply, the accelerometer is specified for a typical noise floor of order 1.4 mg/$\sqrt{Hz}$, a 400 Hz bandwidth, and a 0.038 V/*g* response, yielding an expected noise voltage of order 1 mV rms. We employ a 12 bit A/D measurement with a 5 V full-scale range, yielding 1.2 mV/bit. The A/D converter is specified with a one bit noise floor and is therefore reasonably well matched to the expected noise floor of the device. The data in Fig. 1 indicates a standard deviation of 1.6 bits. Assuming the variances add in quadrature, the resulting noise of the device is of order 1.25 bits, which corresponds to 1.5 mV rms and compares very well with the factory specifications.

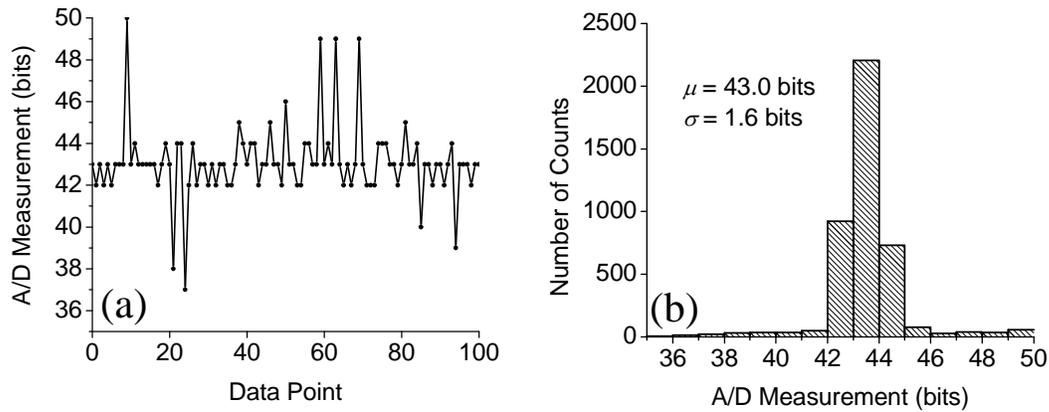

Fig 1. Measurement from the ADXL 78 in the experimental system. The measurement is reported in terms of A/D bits. (a) shows 100 sequential data points, sampled at 4.42 ms per point. (b) is a histogram of a large ensemble of points. The standard deviation of the measurement is 1.6 bits.

Measurement Calibration

The accelerometers are calibrated relative to gravity. Shown in Fig. 2 are data obtained from the system with the accelerometers aligned parallel and anti-parallel to the gravitational acceleration of the earth. The average of the two orientations yields the neutral condition (i.e. 0 $g$) and the difference between the two orientations yields the signal associated with 2 $g$ acceleration. The 2 $g$ signal allows the sensors to be calibrated. The resulting calibration yields $18.3 \pm 0.5$ mV/$g$ for the 120 $g$ accelerometer and $39.0 \pm 1.0$ mV/$g$ for the 50 $g$ accelerometer, consistent with the specifications in the data sheets ($18.0 \pm 0.9$ and $38 \pm 1.9$ mV/$g$, respectively).

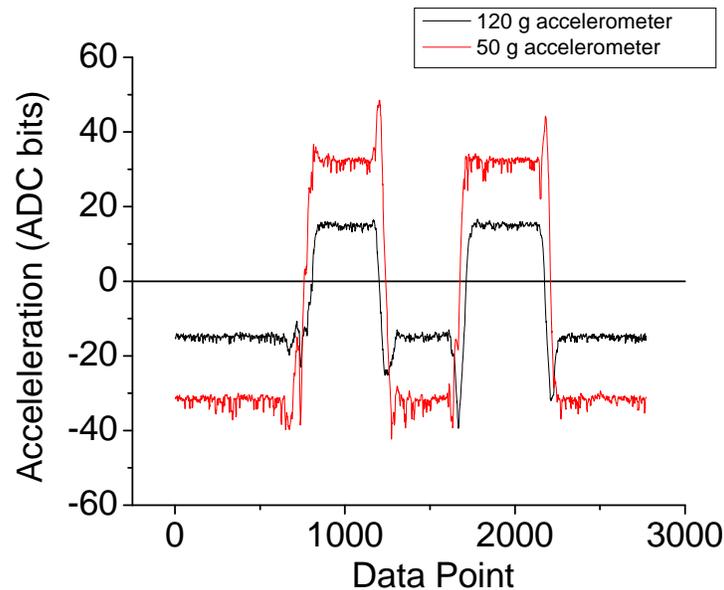

Fig. 2 Calibration of the two sensors against gravity. The club is switched between two positions, aligned parallel and anti-parallel with earth gravity. The ordinate is each successive data point, taken at 4.42 ms/point. The abscissa is the output of the 12 bit A/D converter, calibrated such that the zero-$g$ signal is zero. Note that the 50 $g$ accelerometer has a larger response than does the 120 $g$ accelerometer.

Measurements on a Golf Swing

Shown in Fig. 3 is raw data, $S_1$ and $S_2$, for a single golf swing. $S_1$ is the 120 $g$ accelerometer located near the club head end of the shaft. $S_2$ is the 50 $g$ accelerometer located under the grip of the shaft. The data was taken while swinging a club, but not hitting a ball. This is done so as not to complicate this descriptive analysis with the shock of impact. The data consist of 600 pairs of points, each pair taken at 4.42 ms intervals. The zero of the time axis is arbitrary and corresponds to a point between the beginning of the downswing and impact. The data has been normalized such that the y-axis is calibrated in units of gravitational acceleration, $g = 9.8$ m/s$^2$. The data in (a) and (b) are identical, with (b) being scaled so that one can see the details at small accelerations.

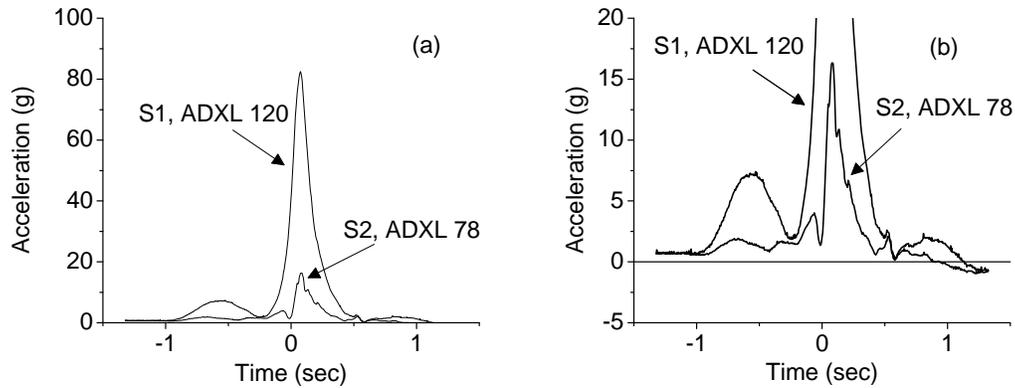

Fig. 3: Raw data from both accelerometers during the golf swing. $S_1$ is the 120 $g$ accelerometer located near the club head end of the shaft. $S_2$ is the 50 $g$ accelerometer located under the grip of the shaft. Graphs (a) and (b) are identical data sets. The y-axis is zoomed in (b) so that one can see the details at low accelerations. The zero of the time axis is arbitrary and corresponds to a moment between the beginning of the downswing and impact.

Quantitative Analysis In Terms of the Rotation of a Rigid Rod

Quantitative analysis of the raw data in Fig. 3 begins by considering the generalized two-dimensional motion of a point on a rigid rod, the geometry of which is shown in Fig. 4. The two dimensional analysis is particularly relevant as it is widely acknowledged that the golf swing occurs in a plane. Everything is referenced to a fixed, inertial, Cartesian coordinate system in the plane of the golf swing with the $\hat{x}$-axis aligned along the direction of gravitational acceleration. The position of the rod in space is defined by the coordinates $\vec{R}_0 = (X_0, Y_0)$ of the reference point $\vec{R}_0$ on the rod and the angle $\phi$ of the rod with respect to the $\hat{x}$-axis. The choice of the point $\vec{R}_0$ is arbitrary, though it is convenient to choose either the center of mass of the rod or a point around

which the rod rotates. The distance to the general point $\vec{r}_1$ on the shaft is measured relative to the reference point $\vec{R}_0$. The coordinates of $\vec{r}_1$ are given as

$$\vec{r}_1 = (X_0 + r_1 \cos\phi)\hat{x} + (Y_0 + r_1 \sin\phi)\hat{y} \tag{1}$$

Taking two time derivatives, one expresses the acceleration of the point $\vec{r}_1$ as

$$\ddot{\vec{r}}_1 = \left(\ddot{X}_0 - r_1\dot{\phi}^2 \cos\phi - r_1\ddot{\phi}\sin\phi\right)\hat{x} + \left(\ddot{Y}_0 - r_1\dot{\phi}^2 \sin\phi + r_1\ddot{\phi}\cos\phi\right)\hat{y} \tag{2}$$

It is useful to rewrite this equation in terms of the in terms of the $\hat{r}$-$\hat{\phi}$ coordinate system, as indicated in Fig. 4. Using the relations

$$\hat{x} = \hat{r}\cos\phi - \hat{\phi}\sin\phi \tag{3a}$$

$$\hat{y} = \hat{r}\sin\phi + \hat{\phi}\cos\phi \tag{3b}$$

one obtains

$$\ddot{\vec{r}}_1 = \left(\ddot{X}_0 \cos\phi + \ddot{Y}_0 \sin\phi - r_1\dot{\phi}^2\right)\hat{r} + \left(-\ddot{X}_0 \sin\phi + \ddot{Y}_0 \cos\phi + r_1\ddot{\phi}\right)\hat{\phi} \tag{4}$$

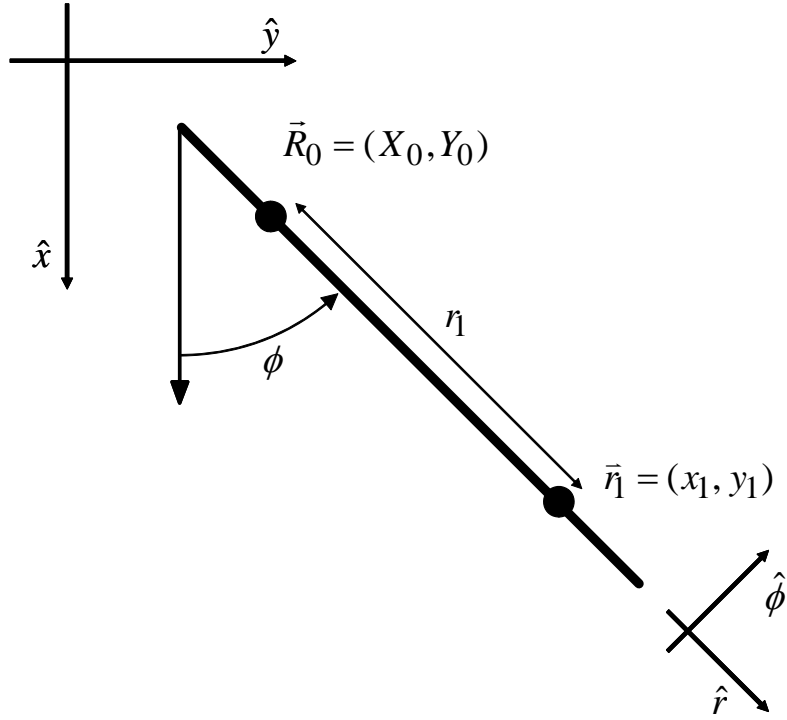

Fig. 4: The geometry of the motion of a rigid rod in a fixed plane. The various coordinates are defined in the text.

Now consider the case of two accelerometers oriented along the shaft at positions $\vec{r}_1$ and $\vec{r}_2$ measured relative to the reference point on the rod, $\vec{R}_0$. Because the accelerometers measure acceleration along the axis of the shaft oriented so as to yield a positive centripetal acceleration, one projects the acceleration along the negative $\hat{r}$-axis, yielding:

$$S_1 = -\hat{r} \cdot \ddot{\vec{r}}_1 = -\ddot{X}_0 \cos\phi - \ddot{Y}_0 \sin\phi + r_1 \dot{\phi}^2 \tag{5a}$$

$$S_2 = -\hat{r} \cdot \ddot{\vec{r}}_2 = -\ddot{X}_0 \cos\phi - \ddot{Y}_0 \sin\phi + r_2 \dot{\phi}^2 \tag{5b}$$

It happens that these measurements are made in the presence of earth's gravitational field. The above equations can be adjusted to include this effect, yielding the expressions

$$S_1 = -\hat{r} \cdot \ddot{\vec{r}}_1 = -\ddot{X}_0 \cos\phi - \ddot{Y}_0 \sin\phi + r_1\dot{\phi}^2 + g^* \cos\phi \qquad (6a)$$

$$S_2 = -\hat{r} \cdot \ddot{\vec{r}}_2 = -\ddot{X}_0 \cos\phi - \ddot{Y}_0 \sin\phi + r_2\dot{\phi}^2 + g^* \cos\phi \qquad (6b)$$

where $g^*$ is the effective gravitational acceleration in the plane of the golf swing.

These two signals can be separated into two components. The first is a common mode, $F(t) = -\ddot{X}_0 \cos\phi - \ddot{Y}_0 \sin\phi + g^* \cos\phi$, and the second is a differential mode $G(t) = (r_1 - r_2)\dot{\phi}^2$. Thus, we can rewrite $S_1$ and $S_2$ in the generic formalism

$$S_1 = F(t) + \frac{r_1}{r_1 - r_2} G(t) \qquad (7a)$$

$$S_2 = F(t) + \frac{r_2}{r_1 - r_2} G(t) \qquad (7b)$$

The differential mode signal, $G(t)$, is recovered by taking the difference of the two signals, $S_1 - S_2 = G(t) = (r_1 - r_2)\dot{\phi}^2$. Because the separation between the two accelerometers $r_1$-$r_2$ is easily measured, determining $G(t)$ enables calculation of $\phi(t)$. The differential mode measurement is proportional to the rotational kinetic energy of the rod, $E_{rot} = I\dot{\phi}^2$, where $I$ is the moment of inertia of the rod about its center of rotation. This positive definite signal is sufficiently simple to generate and sufficiently easy to interpret that it provides a robust foundation on which to base a mechanism for real-time biofeedback for the golfer [3] on issues related to the speed of the club: tempo, rhythm, and timing, in particular.

While the differential mode signal is independent of the choice of the point $\vec{R}_0$, the common mode signal depends strongly on the choice of the point $\vec{R}_0$, which is arbitrary. For example, translating the point $\vec{R}_0 = (X_0, Y_0)$ along the rod through a

distance $\Delta r$, yields $\vec{R}_0^* = (X_0^*, Y_0^*) = (X_0 + \Delta r \cos\phi, Y_0 + \Delta r \sin\phi)$. The resulting

common mode signal $F^*(t) = -\ddot{X}_0^* \cos\phi - \ddot{Y}_0^* \sin\phi$ can be written in terms of the

original common mode signal $F^*(t) = F(t) + \dfrac{\Delta r}{r_1 - r_2} G(t)$. Thus, the choice of the point

$\vec{R}_0$ in any analysis determines how much of the differential mode signal is mixed into the

calculated common mode signal. This ambiguity makes recovering $F(t)$ is a bit more

difficult. In particular, it requires a model for the motion of the point $\vec{R}_0 = (X_0, Y_0)$. To

this end we employ the simplest quantitative model of the golf swing: the double

pendulum.

Quantitative Analysis In Terms of the Double Pendulum

      Use of the double pendulum for analysis of the golf swing was developed by T.P.

Jorgensen [2]. The model is shown schematically in Fig. 5. Our implementation

assumes no translational motion of the center of the swing and it assumes all motion is

confined to a plane. Additionally, it assumes a rigid shaft, which is reasonable given that

shaft dynamics are a secondary effect [4]. The link between the club and the body is

represented as a rigid rod of length $l_0$ oriented at an angle $\theta$ with respect to the *x*-axis of

the inertial, Cartesian, coordinate system fixed in the plane of the swing with the x-axis

aligned along the direction of gravitational acceleration. The golf club is modeled as a

rigid rod of length $l_c$ at an angle $\phi$ with respect to the *x*-axis. The orientation of the golf

club is traditionally measured by the angle $\beta = \theta - \phi$, which roughly corresponds to the

angle through which the wrists are cocked. Note here that $\beta$ is measured in the opposite orientation from $\theta$ and $\phi$ and is consistent with the definition of Jorgensen.

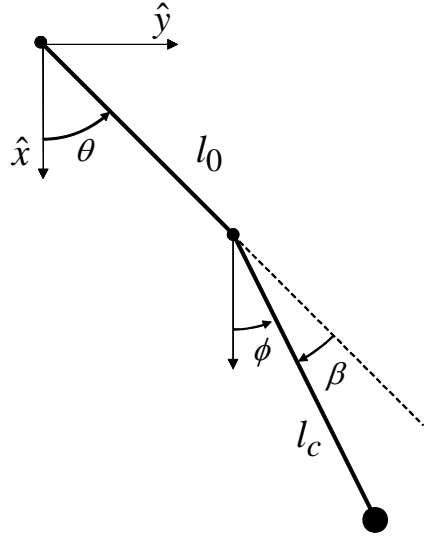

Fig 5. Geometry of the double pendulum. The angle $\theta$ defines the angle of the upper arm, $l_0$ with respect to the x-axis and defines the angle of the lower arm, $l_c$, with respect to the x-axis. The angle $\beta$ defines the angle of the lower arm with respect to the upper arm, and is interpreted as the wrist cocking angle.

The accelerometers are oriented along the axis of the golf club, their positions along the club measured from the hinged coupling of the two rods and given by the lengths $r_1$ and $r_2$. Following as in the analysis above, their position in space is given as

$$\vec{r}_1 = (l_0 \cos\theta + r_1 \cos\phi)\hat{x} + (l_0 \sin\theta + r_1 \sin\phi)\hat{y} \tag{8a}$$

$$\vec{r}_2 = (l_0 \cos\theta + r_2 \cos\phi)\hat{x} + (l_0 \sin\theta + r_2 \sin\phi)\hat{y} \tag{8b}$$

One can determine the generalized acceleration of the two points $\vec{r}_1$ and $\vec{r}_2$ as

$$\ddot{\vec{r}}_1 = -\left(l_0\dot{\theta}^2 \cos\theta + l_0\ddot{\theta}\sin\theta + r_1\dot{\phi}^2 \cos\phi + r_1\ddot{\phi}\sin\phi\right)\hat{x}$$
$$- \left(l_0\dot{\theta}^2 \sin\theta - l_0\ddot{\theta}\cos\theta + r_1\dot{\phi}^2 \sin\phi - r_1\ddot{\phi}\cos\phi\right)\hat{y} \tag{9a}$$

$$\ddot{\vec{r}}_2 = -\left(l_0\dot{\theta}^2 \cos\theta + l_0\ddot{\theta}\sin\theta + r_2\dot{\phi}^2 \cos\phi + r_2\ddot{\phi}\sin\phi\right)\hat{x}$$
$$-\left(l_0\dot{\theta}^2 \sin\theta - l_0\ddot{\theta}\cos\theta + r_2\dot{\phi}^2 \sin\phi - r_2\ddot{\phi}\cos\phi\right)\hat{y} \quad (9b)$$

It is useful to rewrite the above the equations in terms of the $r$-$\phi$ coordinate system attached to the golf club with the $r$-axis aligned along the shaft. Using the relations

$$\hat{x} = \hat{r}\cos\phi - \hat{\phi}\sin\phi \quad (10a)$$

$$\hat{y} = \hat{r}\sin\phi + \hat{\phi}\cos\phi \quad (10b)$$

and the trigonometric identities

$$\sin\theta\cos\phi - \cos\theta\sin\phi = \sin(\theta - \phi) \quad (11a)$$

$$\sin\theta\sin\phi + \cos\theta\cos\phi = \cos(\theta - \phi) \quad (11b)$$

one obtains

$$\ddot{\vec{r}}_1 = -\left(r_1\dot{\phi}^2 + l_0\dot{\theta}^2 \cos\beta + l_0\ddot{\theta}\sin\beta\right)\hat{r} + \left(r_1\ddot{\phi} - l_0\dot{\theta}^2 \sin\beta + l_0\ddot{\theta}\cos\beta\right)\hat{\phi} \quad (12a)$$

$$\ddot{\vec{r}}_2 = -\left(r_2\dot{\phi}^2 + l_0\dot{\theta}^2 \cos\beta + l_0\ddot{\theta}\sin\beta\right)\hat{r} + \left(r_2\ddot{\phi} - l_0\dot{\theta}^2 \sin\beta + l_0\ddot{\theta}\cos\beta\right)\hat{\phi} \quad (12b)$$

Projecting the acceleration along the negative $\hat{r}$-axis yields a positive centripetal acceleration:

$$S_1 = -\hat{r} \cdot \ddot{\vec{r}}_1 = r_1\dot{\phi}^2 + l_0\dot{\theta}^2 \cos\beta + l_0\ddot{\theta}\sin\beta + g^*\cos\phi \quad (13a)$$

$$S_2 = -\hat{r} \cdot \ddot{\vec{r}}_2 = r_2\dot{\phi}^2 + l_0\dot{\theta}^2 \cos\beta + l_0\ddot{\theta}\sin\beta + g^*\cos\phi \quad (13b)$$

Gravitational acceleration has been added to these equations and note that the magnitude of $g^*$ is the projection of the gravitational acceleration into the plane of motion.

Consistent with the earlier results, the common mode and differential mode signals are given as

$$G(t) = (r_1 - r_2)\dot{\phi}^2 \quad (14a)$$

$$F(t) = l_0\left(\dot{\theta}^2 \cos\beta + \ddot{\theta} \sin\beta\right) + g^* \cos\phi \qquad (14b)$$

where we now have replaced the generic terms $\ddot{X}_0$ and $\ddot{Y}_0$ with explicit expressions in terms of the motion of the double pendulum. This allows us to write the two signals as before,

$$S_1 = F(t) + \frac{r_1}{r_1 - r_2} G(t) \qquad (15a)$$

$$S_2 = F(t) + \frac{r_2}{r_1 - r_2} G(t) \qquad (15b)$$

Signal Analysis

One goal for signal analysis is to use the measurements $S_1$ and $S_2$ to determine $\theta(t)$ and $\phi(t)$. In the following, we discuss the issues associated with solving this problem.

One obtains $G(t)$ simply by taking the difference $S_1 - S_2$. $\phi(t)$ is calculated from $G(t)$ as

$$\dot{\phi} = \pm \sqrt{\frac{S_1 - S_2}{r_1 - r_2}}. \qquad (16)$$

It is straightforward to measure the separation between sensors, $r_1 - r_2$. The sign convention is negative in the backswing and positive in the downswing. $\dot{\phi}(t)$ is integrated to yield $\phi(t)$ provided the initial conditions $\phi_i$ and $\dot{\phi}_i$ are known. The initial condition $\dot{\phi}_i = 0$ is accurate just before the beginning of the swing; however, $\phi_i$ is ambiguous. $\phi_i$ can either be measured directly before the swing or after the swing using

video analysis. Generically, $\phi_i$ is constrained relatively close to zero, usually between 5 and 20 degrees.

Determining *F(t)* from $S_1$ and $S_2$ is indefinite. The problem is that we do not know and can not measure $r_1$ and $r_2$, as we do not know exactly around which point the club rotates. It is reasonable to assume that this point is between the hands, but exactly where the golfer grips the club can vary from shot to shot and locating this point somewhere within the hands introduces error of order 10-15% due to the spatial extent of the grip. Thus, it would be nice to identify a technique that allows the determination of *F(t)* without any prior knowledge of the exact values of $r_1$ and $r_2$. The following algorithm is suggested as a reasonable solution.

Assume the function $S_i(t)$ of the form $S_i(t) = F(t) + a_i G(t)$ where one knows *G(t)* but does not know either $a_i$ or *F(t)*. One strategy for determining $a_i$ is to minimize the quantity $\int dt [S_i(t) - a_i G(t)]^2$, which is equivalent to assuming *F(t)* and *G(t)* are orthogonal functions, $\int dt F(t) G(t) = 0$. Taking a derivative with respect to $a_i$ yields the expression

$$a_i = \frac{\int dt\, S_i(t) G(t)}{\int dt\, G(t)^2}.$$

Using this expression for $a_i$, the common mode is calculated as $F(t) = S_i(t) - a_i G(t)$. Additionally, the lengths $r_i$ are discovered, $r_i = a_i (r_1 - r_2)$.

Figures 6(a) and 6(b) display the result of this calculation using the data set shown in Fig 3. The data in Fig. 6(a) is the result for *G(t)* and the data in Fig. 6(b) is the result for *F(t)*. The acceleration is normalized relative to *g*, the gravitational acceleration at

the surface of the earth. A detailed, comparative study of these signals for different golfers is the subject of a second paper [5]. The interpretation is summarized as follows:

1) The function $G(t) = (r_1 - r_2)\dot{\phi}^2$ is a reasonable proxy for the speed of the club. From $G(t)$ many details about the tempo, rhythm, and timing of the swing can be determined, such as the duration of the backswing and downswing. 2) The function $F(t)$ yields information about the motion of the point about which the club is rotating, $R_0$. In the golf swing, this is essentially the motion of the hands. As was detailed above, $F(t) = l_0(\dot{\theta}^2 \cos\beta + \ddot{\theta}\sin\beta) + g^* \cos\phi$. By definition, the magnitude of $g^*$ is less than unity, and is thus not a dominant term. For most of the backswing and downswing, the hands are cocked and thus $\beta$ is of order $\pi/2$. Thus $F(t) \sim l_0\ddot{\theta}$, which is a measure of the acceleration of the hands.

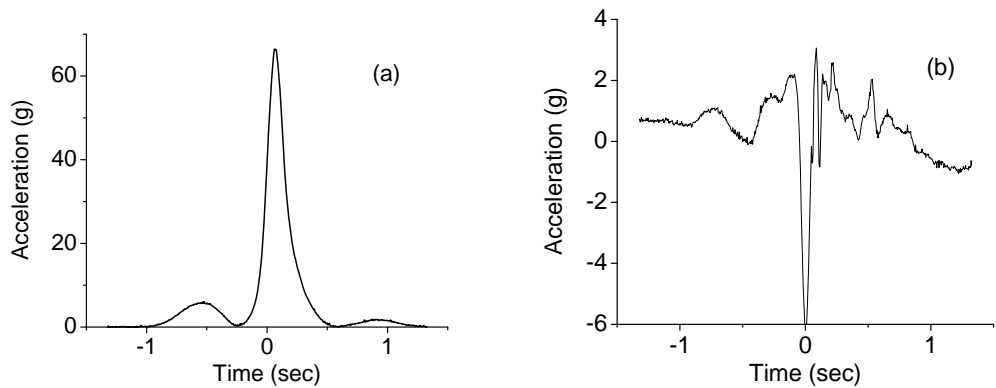

Fig 6. (a) is $G(t)$ and (b) is $F(t)$ calculated from the data shown in Fig 3, above. The calculation follows the algorithm discussed in the text. The acceleration is normalized relative to 9.8 m/s$^2$, the gravitational acceleration at the surface of the earth.

Experimental Validation of $F(t)$

There is no *a priori* reason to assume orthogonality of the common mode and differential mode signals based on the expressions given in Eqs 14a and 14b. In fact, one

might expect this algorithm to yield errors in the determination of $a_1$ and $a_2$, which are equivalent to errors in the distances $r_1$ and $r_2$. Thus, a reasonable test of this algorithm is to perform the above calculation for several golf swings, comparing the resulting values of $r_1$ and $r_2$. to the position of the hands on the club. If our algorithm is correct, it should turn out that the point of rotation $R_0$ is approximately at the midpoint of the grip, between the left and right hands.

      To validate our algorithm, measurements of the golf swing were performed as a function of the position of the placement of the hands on the golf club. The experimental geometry is described as follows. Sensor $S_1$ is located 28.25 inches from the butt end of the golf club, while sensor $S_2$ is located 6.75 inches from the butt end of the golf club, yielding $r_1$-$r_2$ = 21.5 inches. The hands of the right handed golfer are positioned on the golf club with the position of the pinky finger of the left hand initially at the butt end of the club, and then moved down the shaft in 2" increments. Five swings of the club are recorded for each position of the hands. Values of $r_1$ and $r_2$ are calculated for each swing according to the algorithm described above, the averages of which are shown in Fig. 7. The linear fit to the data for both $r_1$ and $r_2$ yields a slope equal to $0.98 \pm 0.01$, consistent with the expected slope of one. The position of the point $R_0$ is consistently located approximately half way between the left and right hands, 2.2 inches from the left pinky finger of the right handed golfer. This is just about exactly where the left fore finger and the right pinky finger overlap when using the classic Vardon grip. It is perfectly reasonable to assume the club rotates about a point in this location, thereby validating the data analysis protocol described above.

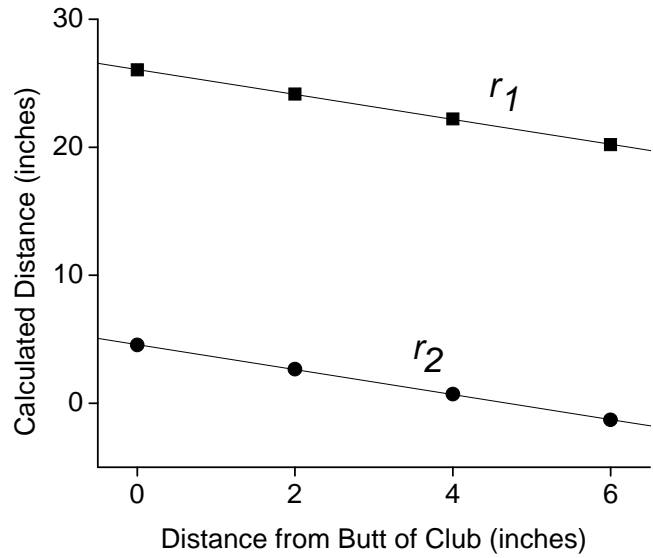

Figure 7: The calculated values of $r_1$ and $r_2$ as a function of the distance of the left pinky from the butt end of the club. The solid lines are a linear fit to the data and in both cases yield a slope equal to $0.98 \pm 0.01$.

Signal Analysis, Continued

Once $F(t)$ is determined, Eq. 14b can be used to solve for $\theta(t)$. The value of $g^*$ is determined from the value of $F(t)$ just prior to the beginning of the swing, when $\dot{\theta}(t)$ and $\ddot{\theta}(t)$ are assumed to be zero and $\phi_i$ is known. Having previously determined $\phi(t)$ from $G(t)$, one can now reliably subtract $g^* \cos\phi$ from $F(t)$, yielding

$$\xi(t) = l_0\left(\dot{\theta}^2 \cos\beta + \ddot{\theta} \sin\beta\right) \qquad (17)$$

In principle, Eq. 17 can be used to solve for $\theta(t)$. One particular approach is to seek an update equation. Assume that at time $t$, one knows $\theta(t)$, $\dot{\theta}(t)$, and $\ddot{\theta}(t)$. Then assume the following:

$$\ddot{\theta}(t + dt) = \ddot{\theta}(t) + \varepsilon \qquad (18a)$$

$$\dot{\theta}(t+dt) = \dot{\theta}(t) + \left(\frac{\ddot{\theta}(t+dt) + \ddot{\theta}(t)}{2}\right) dt \qquad (18b)$$

$$\theta(t+dt) = \theta(t) + \left(\frac{\dot{\theta}(t+dt) + \dot{\theta}(t)}{2}\right) dt \qquad (18c)$$

Defining the following constants

$$\dot{\theta}_0 = \dot{\theta}(t) + dt\,\ddot{\theta}(t) \qquad (19a)$$

$$\theta_0 = \theta(t) + dt\,\dot{\theta}(t) + \frac{dt^2}{2}\ddot{\theta}(t) \qquad (19b)$$

and expanding Eqs 18(a), 18(b), and 18(c) above, one finds the expressions

$$\dot{\theta}(t+dt) = \dot{\theta}_0 + dt\,\frac{\varepsilon}{2} \qquad (20a)$$

$$\theta(t+dt) = \theta_0 + dt^2\,\frac{\varepsilon}{4} \qquad (20b)$$

Inserting these expressions into Eq 17 above, one obtains

$$\frac{\xi(t+dt)}{l_0} = (\ddot{\theta}_0 + \varepsilon)\sin\left(\beta_0 + \varepsilon\frac{dt^2}{4}\right) + \left(\dot{\theta}_0 + \varepsilon\frac{dt}{2}\right)^2 \cos\left(\beta_0 + \varepsilon\frac{dt^2}{4}\right) \qquad (21)$$

where we have defined $\beta_0 = \theta_0 - \phi(t+dt)$. Expanding the above equation to first order in $\varepsilon$ yields the expression

$$\varepsilon = \frac{\dfrac{\xi(t+dt)}{l_0} - \ddot{\theta}_0 \sin\beta_0 - \dot{\theta}_0^2 \cos\beta_0}{\sin\beta_0 + dt\,\dot{\theta}_0 \cos\beta_0 + \dfrac{dt^2}{4}\left(\ddot{\theta}_0 \cos\beta_0 - \dot{\theta}_0^2 \sin\beta_0\right)} \qquad (22)$$

Thus, Eq. 22 can be used as an update equation to determine $\theta(t+dt)$, $\dot{\theta}(t+dt)$, and $\ddot{\theta}(t+dt)$. This can be solved to higher order in $\varepsilon$, if increased numerical precision is deemed necessary. The only parameter in this update formula that can not be precisely

measured is $l_0$. However, $l_0$ is constrained within relative limits of order 10% if one knows the geometry of the particular golfer.

Implementing the Signal Analysis

The goal of the signal analysis is to solve for $\phi(t)$ and $\theta(t)$. We have done this for the data shown in Fig. 6. We only perform this analysis up to impact, as in real golf swings the shock associated with impact yields electrical transients that do not allow us to integrate past impact.

Calculating $\phi(t)$ is straightforward, requiring one to integrate Eq. 16. Using this result in combination with Eq. 18 and Eq. 22 to calculate $\theta(t)$ is a little more difficult. First, there are three adjustable parameters in this analysis: the length of the upper arm of the pendulum, $l_0$; the angle at address, $\beta_i$; and the angle at impact, $\beta_f$. Our strategy for solving the problem is to start by fixing $\beta_i$ and integrating the equations of motion, adjusting $l_0$ so as to obtain a solution that terminates at $\beta_f$. As will be shown below, this calculation is relatively insensitive to the choice of $\beta_f$ and unless otherwise stated, we will assume $\beta_f = \beta_i$. We iterate through several such calculations, systematically varying the initial angle $\beta_i$ until we find those initial conditions that are consistent with the physical bounds of $l_0$.

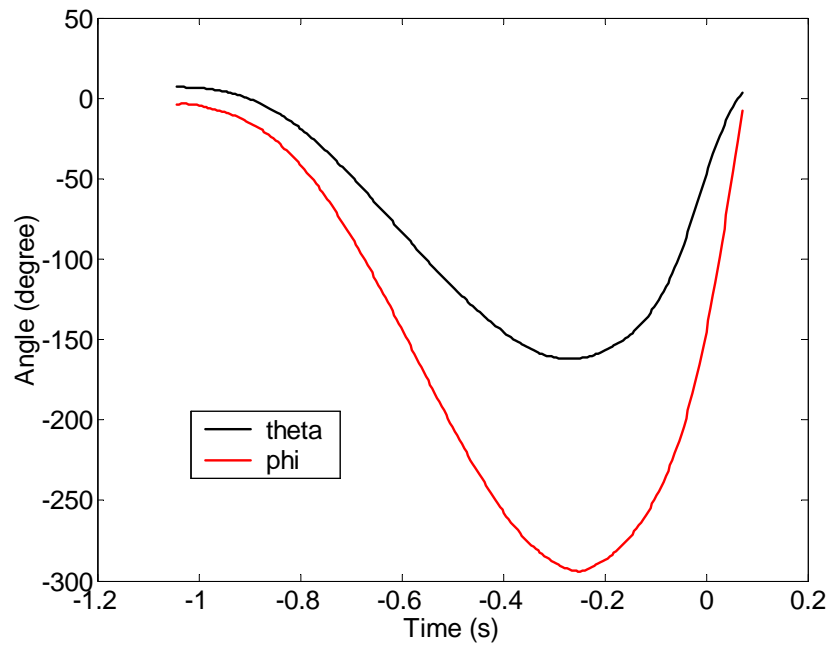

Figure 8: The calculated values of $\phi(t)$ and $\theta(t)$ as a function of time. The zero of time is arbitrary. The initial and final angles $\beta_i = \beta_f = 10.5$ degrees. The resulting value of $l_0$ is 0.48 meters. The maximum club head speed at impact was calculated to be 84 miles per hour.

Shown in Fig. 8 are the calculated values of $\phi(t)$ and $\theta(t)$ for the conditions $\beta_i = \beta_f = 10.5$ degrees. The resulting value of $l_0$ is 0.48 meters, consistent with the estimate for our model golfer of $0.5 \pm 0.05$ meters. The maximum speed at impact was calculated to be 84 mph, consistent with our external measurement of 82 mph [6].

This calculation has been performed as a function of the initial angle $\beta_i$. A summary of these results are shown in Fig 9. The initial angle $\beta_i$ spans 10-25 degrees. The resulting values of $l_0$ vary from 0.60 to 0.13 m. The actual value of $l_0$ for the golfer under study was $0.5 \pm 0.05$. Thus, our approximate knowledge of $l_0$ limits $\beta_i$ to the very narrow range $10.5 \pm 0.5$ degrees.

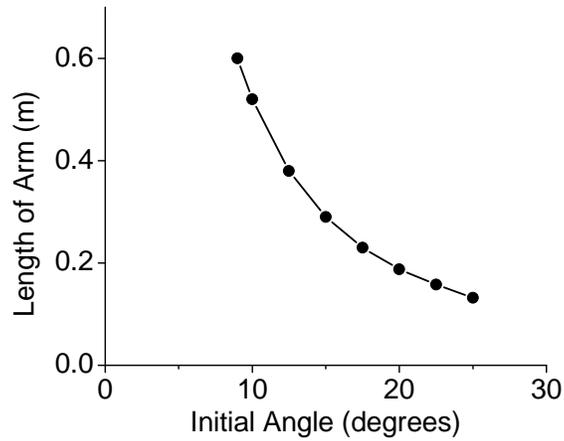

Figure 9: The length $l_0$ as a function of initial angle $\beta_i$, as determined by finding that value of $l_0$ which yields and end point $\beta_f = \beta_i$. The value of $l_0$ is very insensitive to the exact choice of $\beta_f$.

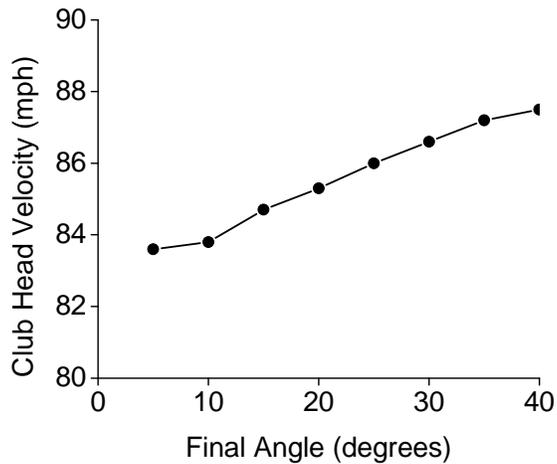

Figure 10: The club head velocity as a function of the final angle $\beta_f$ at impact. The initial angle $\beta_i$ was set to 10.5 degrees. In each case, the resulting value of $l_0$ was 0.486 m. This calculation shows how insensitive the calculation is to on the final angle, $\beta_f$.

In an effort to quantify how insensitive the calculation is to the choice of $\beta_f$, the above calculations have been performed as a function of $\beta_f$. These calculations keep

$\beta_i$ constant at 10.5 degrees, as determined above. The final value $\beta_f$ was varied from 5 to 40 degrees. In each case, the calculated value of $l_0$ turned out to be 0.48 m, indicating the insensitivity of $l_0$ on the choice of $\beta_f$. Shown in Fig. 10 is the calculated club head speed at impact as a function of $\beta_f$. Throughout the entire range of this calculation, the resulting club head speed varies by no more than 2 mph from the average value of 85.5 mph. Thus, we contend that the entire calculation is relatively insensitive to the value of $\beta_f$, justifying our choice $\beta_f = \beta_i$.

Finally, the orientation of the double pendulum as a function of time was calculated for the data in Fig. 8. The result is shown in Fig 11(a) for the backswing and Fig. 11(b) for the downswing.

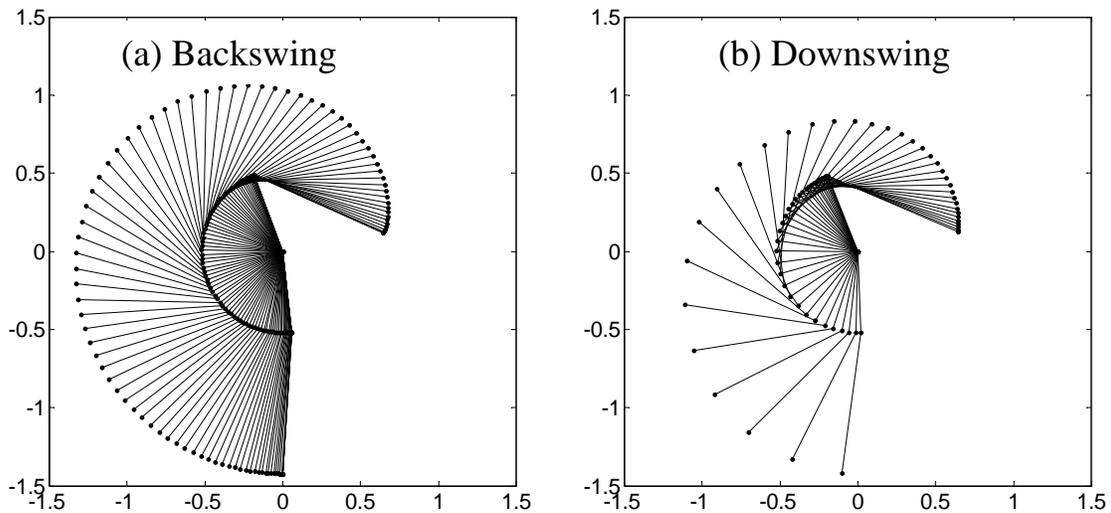

Fig 11: Resulting motion of the double pendulum, as calculated from the data in Fig 8 above. The image in (a) is the backswing and the image in (b) represents the downswing. For clarity, the images show only every other data point. The axes are calibrated in units of meters.

Conclusion

In summary, we have described a measurement system that mounts in the shaft of a golf club. The measurement requires two accelerometers, each placed at opposite ends of the golf shaft with the sensing axis oriented along the shaft of the golf club.

It is shown that this relatively simple measurement technique can be interpreted within the context of the double pendulum model of the golf swing. This model assumes the golf swing rotates around a fixed hub and is confined to a plane. The measurement yields sufficient fidelity to calculate the angles $\phi(t)$ and $\theta(t)$, which characterize the orientation of the two arms of the double pendulum in the plane of the swing throughout the entire golf swing.

Footnotes and References: